\documentclass{INTERSPEECH2023}

\usepackage{multirow}
\usepackage{mathrsfs}
\usepackage{booktabs}
\pdfoutput=1

\interspeechcameraready

\title{Spot keywords from very noisy and mixed speech}
\name{Ying Shi$^1$, Dong Wang$^{2*}$, Lantian Li$^3$, Jiqing Han$^{1*}$, Shi Yin$^4$
 \thanks{This work was supported by the National Natural Science Foundation of China (NSFC) under the project No.~62171250.
 D.W. (wangdong99@mails.tsinghua.edu.cn) and J.H. (jqhan@hit.edu.cn) are the corresponding authors.
}
}
\address{
  $^1$School of Computer Science and Technology, Harbin Institute of Technology, China \\
  $^2$Center for Speech and Language Technologies, BNRist, Tsinghua University, China  \\
  $^3$School of Artificial Intelligence, Beijing University of Posts and Telecommunications, China \\
  $^4$Huawei Technologies Co., Ltd.
}
\email{$^{2*}$wangdong99@mails.tsinghua.edu.cn, $^{1*}$jqhan@hit.edu.cn}

\begin{document}

\maketitle

\begin{abstract}
Most existing keyword spotting research focuses on conditions with slight or moderate noise.
In this paper, we try to tackle a more challenging task: detecting keywords buried under
strong interfering speech (10 times higher than the keyword in amplitude), and even worse,
mixed with other keywords. We propose a novel Mix Training (MT) strategy that
encourages the model to discover low-energy keywords from noisy and mixed speech.
Experiments were conducted with a vanilla CNN and two EfficientNet (B0/B2) architectures.
The results evaluated with the Google Speech Command dataset demonstrated that the proposed mix training
approach is highly effective and outperforms standard data augmentation and mixup training.
\end{abstract}

\noindent\textbf{Index Terms}: Keyword spotting, mixed speech, mix training

\section{Introduction}
Keyword Spotting (KWS) aims to detect one or multiple focused words or phrases from the background signals~\cite{lopez2021deep}. 
Early research of KWS dates back to the 1960s~\cite{teacher1967experimental},
and the pre-SOTA models were based on hidden Markov models (HMMs)~\cite{rose1990hidden,miller2007rapid}.
Since 2014, deep neural networks have emerged as a new backbone~\cite{chen2014small,he2017streaming,audhkhasi2017end,zhao2021end,van2022feature} 
and boosted KWS performance to a level that can support many practical applications~\cite{hoy2018alexa}, 
at least in conditions with low and mild noise~\cite{LossFunctionKWS,MutiScaleRobustKWS,trainablefrontendKWS}.

However, in more challenging situations with intense interfering speech, drastic accuracy degradation is often observed. 
Two typical situations are:  (1) the keyword is significantly weaker than the interfering speech, e.g., only 1/10 in amplitude; 
(2) two or more keywords are mixed together and the keywords corrupt each other. 
As shown in Section~\ref{sec:res}, the top-1 accuracy rates of a model trained with clean speech will drop from 96\% to 16\% in both conditions. 
This is not surprising as detecting weak signals from strong interference is hard, if not possible, even for human ears. 
This is especially true when the interference speech contains keywords, as in the keyword mixing scenario.
Nevertheless, detecting weak keywords is highly valuable and deserves careful investigation.

Data augmentation (DA)~\cite{robustacosticmodeling,effecitveaug} is an approach that can be considered immediately. 
By artificially corrupting the input speech while keeping the label unchanged, the model becomes immune to the same or similar corruptions. 
In this paper, DA specifically refers to mixing with interfering speech.
It has been demonstrated that DA is highly effective in improving model robustness against mild interfering speech such as TV noise~\cite{dataaugkwsplayback,sewakeupword}. 
However, in theory, DA cannot deal with keyword mixing, as such kind of data is not seen during the training process.

Mixup~\cite{mixup} is another promising approach to improve model robustness. 
Different from DA, mixup mixes two keywords by linear interpolation, rather than a keyword and an interference. 
Moreover, the labels of the two keywords are also interpolated to represent the co-existence of two keywords in mixed speech:
\begin{equation}
     \begin{split}
     x' &= \lambda x_i + (1 - \lambda)  x_j  \\
     y' &= \lambda y_i + (1 - \lambda)  y_j.
     \end{split}
     \label{eq:mixup}
\end{equation}
\noindent where $(x_i, y_i)$ and $(x_j,y_j)$ are two training datum randomly sampled from the training set, and $\lambda$ is a random variable sampled from a beta distribution. 

Zhang et al.~\cite{mixup} demonstrated that mixup training leads to improved robustness by minimizing the vicinal risk, rather than empirical risk.  
For our purpose, mixup is a potential approach to spotting mixed keywords, since it shows the model what a keyword mixture looks like. 
However, mixup is also problematic. The interpolated label $y'$ is a compromise of $y_i$ and $y_j$, 
indicating that neither $y_i$ nor $y_j$ definitely exists, which is not consistent with the nature of speech signal and our perception
as we can clearly hear both of them.

\begin{figure*}[htb!]
     \centering
     \includegraphics[width=0.92\linewidth, height=0.32\linewidth]{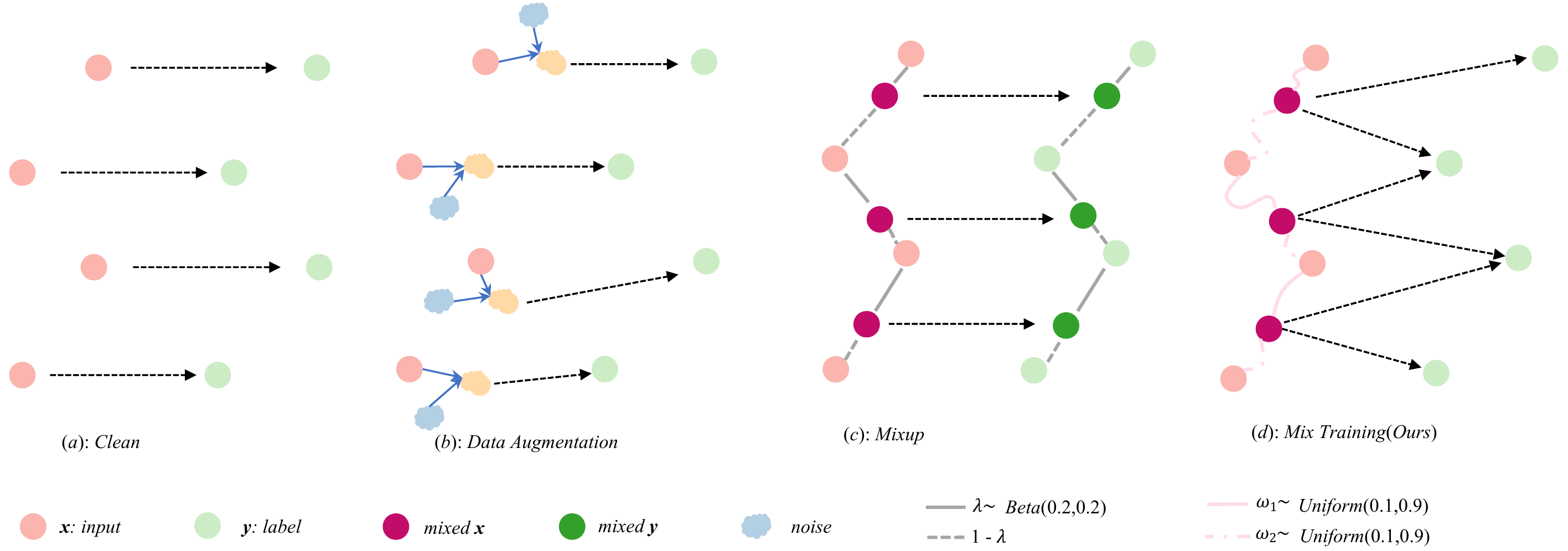}
     \caption{Illustration of mix training and other strategies. (a) Training with clean speech; (b) Data augmentation, where the input data is mixed with interference speech while the label does not change;
     (c) Mixup, where both the input speech and the labels are linearly interpolated. (d) Mix training, where the input speech of two or more keywords are mixed in an arbitrary way, and the labels are the union of the labels of all the component keywords.
     }
     \label{fig:structure}
\end{figure*}

In this paper, we propose a simple but efficient approach called Mix Training (MT). The main operation involves (1) mixing $k$ keywords with arbitrary weights and (2) setting the label to be $k$-hot to 
indicate the definite existence of the $k$ keywords. The method is illustrated in Fig.~\ref{fig:structure} and the details will be presented in Section~\ref{sec:theory}.
At the first glance, this may seems like a minor modification of mixup or trivial extension of DA, but we will show that the change is significant as it reflects the principle of superposition, 
a fundamental property of speech signals, as well as the behavior of human auditory system.

\section{Related work}

Several methods have been developed to improve the robustness of KWS models, such as by ensembling different scale models~\cite{yang2020multi}, pre-training and transfer learning~\cite{park2021noisy,seo2021wav2kws},
and gain control~\cite{wang2017trainable}.

DA has been widely employed in multiple speech processing tasks~\cite{yin2015noisy,effecitveaug}.
For KWS, it is also a standard technique to increase model robustness~\cite{dataefficentmodelkws,depthwisekws,luo2022speech}. 
 Some studies have used DA to tackle interfering speech, e.g., \cite{dataaugkwsplayback,sewakeupword} used TV recordings as noise to 
 augment keyword speech. 

Mixup was first proposed by Zhang et al.~\cite{mixup}. Interestingly, KWS was a demonstration task in this seminal work, 
and the authors showed that mixup training produced SOTA performance on the Google Speech Command database.
Since then, a multitude of mixup variants have been proposed, including Manifold mixup~\cite{verma2019manifold}, CutMix~\cite{yun2019cutmix}, PixMix~\cite{hendrycks2022pixmix} and CoMixup~\cite{comixup}. 
All the variants follow the same idea of synthesizing data by mixing the input and labels with linear interpolation.
Theoretical analysis has revealed that mixup training works by adding a regularization term on the directional derivatives of the function represented by the model~\cite{zhang2021does, mixupe}. 
A recent survey paper~\cite{cao2022survey} provides a comprehensive overview of the mixup approach.

\section{Methods}
\label{sec:theory}

In a nutshell, Mix Training (MT) is a method that aims to force the model to recognize \emph{all} the keywords in mixed speech. It
involves three key components: label union, uniform sampling, and binary cross entropy (BCE) loss.

\vspace{-1.5mm}
\subsection{Label union}


The mixup training shown in Eq.(\ref{eq:mixup}) holds two assumptions: (1) the interpolation of two data samples is also a valid data point, 
and (2) the label is a compromise of the two data samples. 
However, both the two assumptions are unclear, especially for images that are clearly not interpolable in the data space. 
For this reason, the functionality of mixup was a mystery for some time, until recently researchers 
found it plays a role in the regularization of directional derivatives~\cite{zhang2021does,mixupe}. 

Fortunately, this difficulty does not exist for speech signals, as they are naturally mixable as mechanical waves. 
However, the label interpolation is not appropriate as we know that the content of mixed speech can be identified 
by human ears. We therefore slightly modify the mixup training as Eq.(\ref{eq:mt}), which forms the foundation of mix training:

\begin{equation}
     \begin{split}
     x' &= \omega_{1}  x_i + \omega_{2}  x_j \\
     y' &=  y_i \oplus y_j
     \end{split}
     \label{eq:mt}
\end{equation}
\noindent where $\omega_{1}, ~\omega_{2}$ are two independent scale factors, and $\oplus$ denotes logical addition.

Compared to Eq.(\ref{eq:mixup}), there are two significant differences: (1) $x_i$ and $x_j$ can be mixed in any way, not limited to linear interpolation. This 
free mixing corresponds to the superpositional property of speech.
(2) the label $y'$ of the mixed speech represents the `union' of the labels of $x_i$ and $x_j$, reflecting the way human ears behave. 
We therefore establish the theoretical foundation of mix training based on the superposition property of speech signals and the perceptual behavior 
of the human auditory system. 

In practice, the clean speech is also involved in the training process, to ensure the model sees unmixed data. This is 
different from mixup training where clean data has been involved by sampling $\lambda$ close to 0 or 1. 
It also should be mentioned that Eq.(\ref{eq:mt}) can be easily extended to represent mixing more than two speech keywords, though we will 
leave it in future research.

\subsection{Uniform sampling}

We typically use a uniform distribution to sample the scale factors,
which reflects the fact that the content of speech is energy-invariant.
However, caution must be exercised when designing this uniform sampling because values 
that are too close to 0 will result in weak scaled speech, making it illogical to label it as the original keyword. 
This is consistent with human auditory behavior, as we can only hear sounds whose energy exceeds a threshold~\cite{gelfand2017hearing}.
To address this issue, we set the limit of uniform sampling between 0.1 and 0.9. 
Additionally, to avoid clipping, we also conduct a normalization to ensure $\omega_{1} + \omega_{2} =1$. This looks 
similar to the constraint in mixup training, though it is just a practical treatment in MT and not necessary in theory.

Finally, to learn energy-invariance behavior, the volume scaling is also applied to clean speech training and DA training, 
following the same distribution Uniform (0.1, 0.9).
Moreover, the scale normalization is also applied to DA training to prevent signal clipping. 

\subsection{BCE loss}

\begin{figure}[htbp]
  \centering
     \includegraphics[width=0.80\linewidth, height=0.46\linewidth]{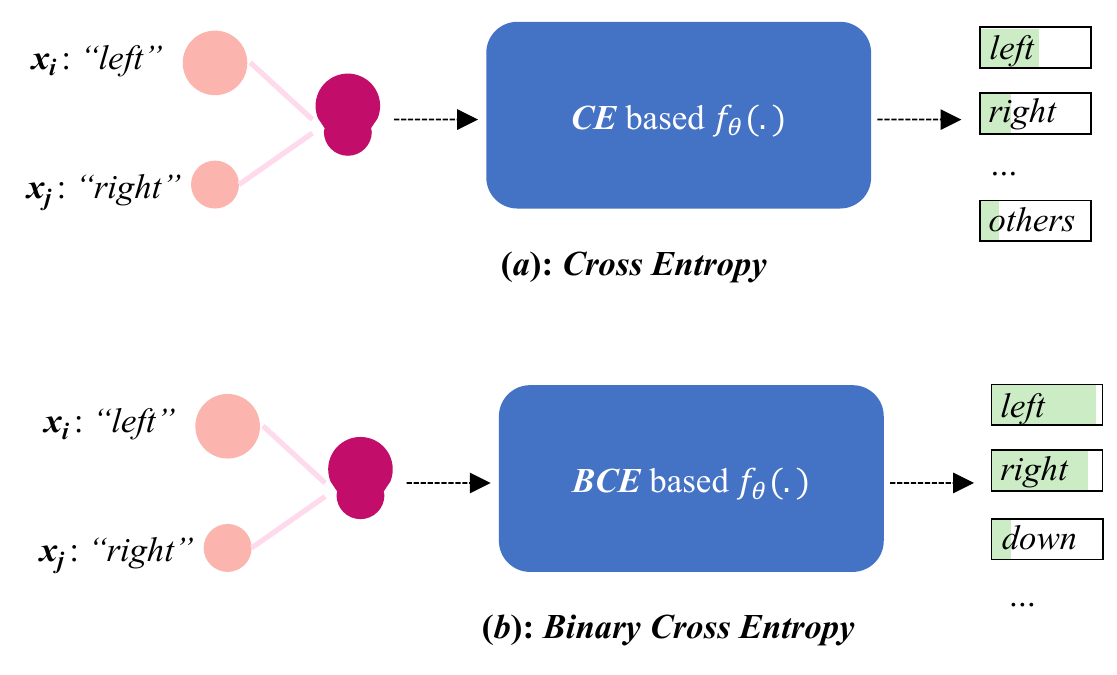}
    \caption{Output of KWS models trained with (a) CE and (b) BCE when the input is a mixture of two keywords. For the CE model, the output represents a categorical distribution, so the values of all the output units sum to 1. For the BCE model, the output represents a Bernoulli distribution and the value of each unit is normalized separately. Note that the CE model cannot represent the co-existence of keywords `left' and `right' but the BCE model can do that.}
     \label{fig:arc}
\vspace{-3.0mm}
\end{figure}

Most KWS models are based on the softmax architecture and trained using cross entropy (CE) loss, as shown in Fig.~\ref{fig:arc} (a). 
However, this approach is unsuitable for detecting mixed keywords, since the softmax outputs are competitive and unable to represent the 
fact that two or more keywords co-exist. We therefore adopt the sigmoid architecture and train the model with binary cross entropy (BCE) loss, 
as shown in Fig.~\ref{fig:arc} (b). 
In this architecture, each output unit corresponds to a particular keyword and its activation reflects the probability that the 
keyword exists. Although BCE is optional for DA and mixup, it is mandatory for MT.
Recent research demonstrated that BCE loss can achieve comparable performance as CE loss on face recognition~\cite{wen2021sphereface2}.
We will also show that CE and BCE models perform similarly on clean models shortly.



\section{Experiments}
\label{sec:exp}
\subsection{Data}
The Google speech command (GSC) v2 dataset~\cite{GSCcorpus} is used to perform the test. The dataset contains 105,829 utterances recorded by 2,618 speakers, 
and includes 35 keywords and a background class. Most of the utterances are 1 second long. 
We follow the official split for training, validation and test sets. 
And especially we use the standard test protocol which involves 12 categories (10 keywords with two additional classes `silence' and `unknown').

We also use the Librispeech/train-clean-100h database as the source of interference speech. 
To prevent potential errors with DA, the utterances that contain the keywords in the GSC corpus are removed.
There are about 12k utterances employed to corrupt the GSC speech, 
11k utterances used for training, and 900 utterances used for testing. 
The two sets of utterances have no speaker overlaps.

\subsection{Model}

Three neural models were utilized to test the performance of various training strategies, as detailed below.

\begin{itemize}
\item A vanilla CNN, which contains 7 convolution blocks.
Each block involves a 2D Convolution layer followed by LayerNorm and ReLU. 
The number of output channels for the seven blocks is 32, 64, 128, 64, 128, 256 and 512, respectively, 
and kernel size is all set to 3.
The stride of the first two blocks is (2,1), and (1,1) for the rest blocks.
After the convolution blocks, a linear layer outputs the probabilities corresponding to 36 classes, including 
35 keywords and a background class.
 
\item EfficientNet-B0~\cite{efficientnet}: The backbone of the EfficientNet-B0, followed by a linear transform to obtain 
the probabilities of the keywords plus the background class. 
The online source code is used\footnote[1]{https://github.com/lukemelas/EfficientNet-PyTorch}.

\item  EfficientNet-B2~\cite{efficientnet}: The same as EfficientNet\_B0, except that the backbone is more powerful.
\end{itemize}

\subsection{Implementation details}
\label{sec:exp:imp}

Speech mixing was performed on the waveform.
For DA, the scale factors of keywords speech and interference speech were randomly sampled from Uniform (0.1, 0.9), and 
normalized to avoid clipping. Only 40\% of the data samples underwent DA, to ensure the performance of clean data.
For mixup, the interpolation factor $\lambda$ is sampled from Beta(0.2,0.2), following the convention 
in~\cite{mixupe}. 
For mix training, the scaling factors $\omega_1$
and $\omega_2$ are sampled from Uniform (0.1, 0.9), and normalized to avoid clipping. 
The input feature is 80-dimensional Fbank with 25 ms windows size and 10 ms frame shift.

All the models are trained for 50 epochs, with a batch size of 128. 
We used the Adam optimizer with a learning rate of 1e-3. 
Warm-up was conducted with 10 epochs. The final model was an average of the last 10 checkpoints.

\section{Results}
\label{sec:res}

Performance was evaluated by two metrics: equal error rate (EER), 
a detection metric that represents the equilibrium point of the false detection rate and the missing rate; 
and top-k accuracy, a classification metric to measure if the probability of the true class is in the top-k positions.

\subsection{CE VS BCE}

We first verified that BCE loss was a vital alternative of the CE loss,
so that we can use BCE to implement comparative techniques without overvaluing one or another. 
We only show baseline models trained with clean data. The results shown in Table~\ref{tab:bce} confirmed that the BCE model and the CE model are
comparable. 

\begin{table}[h!]
  \caption{EER\% and Top-1 Accuracy (Acc\%) of models trained with CE and BCE. All the models were trained with clean speech.}
  \label{tab:bce}
  \vspace{-1.5mm}
  \label{tab:clean}
  \centering
  \scalebox{0.8}{
  \begin{tabular}{lcccccc}
  \toprule
    \multirow{3}{*}{Test Set} & \multicolumn{2}{c}{Vanilla CNN}       & \multicolumn{2}{c}{EfficientNet\_B0}    & \multicolumn{2}{c}{EfficientNet\_B2} \\
                                \cmidrule(r){2-3}                       \cmidrule(r){4-5}                       \cmidrule(r){6-7}
                              &   EER           &  Acc                  &   EER            &  Acc               &   EER             &    Acc   \\
        \cmidrule(r){1-1}       \cmidrule(r){2-3}                       \cmidrule(r){4-5}                        \cmidrule(r){6-7}
        CE                    &   \textbf{1.21} &   95.83               &   1.47           &  96.81             &  1.10             &  \textbf{97.32} \\
        BCE                   &   1.52          &   \textbf{96.01}      &\textbf{0.96}     &\textbf{97.03}      &  \textbf{0.96}    &  97.12 \\
  \bottomrule
  \end{tabular}}
  \vspace{-1.0mm}
\end{table}
\vspace{-1.8mm}
\subsection{Mixed keyword detection}
\label{subsec:mixkwdet}

In this experiment, we tested the capability of various training strategies in detecting mixed keywords. 
To construct the test, we randomly mixed two samples from test the set belonging to the $10$ keywords, following the mixing procedure presented in Section~\ref{sec:exp:imp}.
The mixed speech was used to test if a model could detect each of the original keywords. 
Since there are two keywords in each mixture, we report the Top-2 accuracy.
Table~\ref{tab:mix} reports the results with different training strategies. 
It can be seen that the performance of the clean model drops drastically, indicating that it is severely impacted by mixing. 
DA, mixup and MT were all improving the performance. Mixup and MT, which both involving keyword mixtures in the training process, showed more improvement compared to other strategies.
MT offers the best models and shows a clear advantage over mixup training. 
To provide more diverse mixtures for the mixup training,
we change the sampling distribution for the interpolation factor $\lambda$ from Beta (0.2,0.2) to Uniform (0, 1), the one 
used in mix training. The results are reported in the row denoted by Mixup (U). 
It can be observed that mixtures with greater diversity actually enhance the performance of Mixup, 
as demonstrated by the fact that Mixup (U) outperforms standard Mixup under all conditions.
However, Mixup (U) was still worse than MT, which confirms the importance of label union.

 \begin{table}[h!]
  \caption{Performance in EER\% and Top-2 Accuracy (Acc\%) of different training strategies when detecting mixed keywords.}
  \vspace{-1.5mm}
  \label{tab:mix}
  \centering
  \scalebox{0.8}{
  \begin{tabular}{lcccccc}
  \toprule
    \multirow{3}{*}{} & \multicolumn{2}{c}{Vanilla CNN}    & \multicolumn{2}{c}{EfficientNet\_B0} & \multicolumn{2}{c}{EfficientNet\_B2} \\
                                \cmidrule(r){2-3}                    \cmidrule(r){4-5}                    \cmidrule(r){6-7}
                              &   EER           &  Top2 Acc              &   EER          &  Top2 Acc                &   EER           &  Top2   Acc   \\
        \cmidrule(r){1-1}      \cmidrule(r){2-3}                     \cmidrule(r){4-5}                     \cmidrule(r){6-7}
        Clean                 &   22.14         &   52.87           &  24.69         &  60.59              &  24.95          &  61.60 \\
        DA                    &   17.27         &   64.50           &  20.02         &  69.83              &  20.03          &  70.22 \\
        Mixup                 &    7.54         &   81.46           &   4.07         &  88.36              &   3.76          &  89.19 \\
        Mixup (U)             &    6.59         &   83.24           &   3.37         &  89.68              &   2.99          &  90.91 \\
        MT                    &\textbf{5.73}    &\textbf{85.75}     &\textbf{3.11}   &\textbf{90.35}       &\textbf{2.81}    &\textbf{91.23} \\
  \bottomrule
  \end{tabular}}
\end{table}
To further demonstrate the advantage of MT in detecting mixed keywords, we set the scale factors to be 1:10 when composing the keyword mixture and check if the 
highly weak keywords can be detected. 
The results are shown in Table~\ref{tab:mix1-10NoOverlap}. Note that we only focused on the weak keyword, so we set the output of the strong keywords to 0 and compute top-1 accuracy
with the output of the weak keywords.
In this highly challenging scenario, the clean model almost fails (a decision with more than 40\% EER is nearly random), and the advantage of MT over other robust training methods is 
rather remarkable.

\begin{table}[h!]
  \caption{Performance in EER\% and Top1 Accuracy (Acc\%) of different training strategies when detecting weak keywords from keyword mixtures.}
  \vspace{-1.5mm}
  \label{tab:mix1-10NoOverlap}
  \centering
  \scalebox{0.8}{
  \begin{tabular}{lcccccc}
  \toprule
    \multirow{3}{*}{Test Set} & \multicolumn{2}{c}{Vanilla CNN} & \multicolumn{2}{c}{EfficientNet\_B0} & \multicolumn{2}{c}{EfficientNet\_B2} \\
                                \cmidrule(r){2-3}                 \cmidrule(r){4-5}                    \cmidrule(r){6-7}
                              &   EER   &  Top1 Acc             &   EER   &  Top1 Acc                  &   EER    &    Top1 Acc   \\
        \cmidrule(r){1-1}      \cmidrule(r){2-3}                  \cmidrule(r){4-5}                      \cmidrule(r){6-7}
        Clean                 &   42.65 &   10.73               &   45.74 &  16.58                     &  45.85   &  16.81 \\
        DA                    &   37.17 &   16.15               &   39.03 &  24.04                     &  38.72   &  23.83 \\
        Mixup                 &   20.64 &   38.84               &   13.21 &  57.87                     &  12.36   &  60.40 \\
        Mixup (U)             &   18.09 &   40.72               &   10.38 &  61.09                     &   9.35   &  64.44 \\
        MT                    &\textbf{16.30} &\textbf{56.49}   &\textbf{9.08}&\textbf{67.03}          &\textbf{8.24}&\textbf{69.33} \\
  \bottomrule
\end{tabular}}
\end{table}

\subsection{Noisy keyword detection}

In this experiment, we focused on very noisy conditions, where high-energy interfering human speech was present.
We simulated the scenario by mixing audio segments sampled from Librispeech to keyword speech, 
and set the amplitude of the interfering speech 10 times higher than the keyword speech.

The results are shown in Table~\ref{tab:Noisy1-10}. We observed that both DA, Mixup and
MT provided clear performance improvement, and DA performed the best. This is not surprising since
DA involved the most interfering speech during training. However, Mixup and MT do not use 
extra data, which means the improvement they obtained is \emph{free}. 
Compared to Mixup, the improvement with MT is more significant.
And the result about Mixup (U) confirms our assumption again, which has been illustrated in Section~\ref{subsec:mixkwdet}.

Finally, we trained an MT (N) model, which randomly augmented 40\% of the clean speech with interference 
speech as in DA\footnote{Note MT involves two objectives, one with clean speech and one with mixed speech. The augmentation 
of MT (N) performs augmentation on the clean speech.}. 
This new model obtains significantly better performance than the MT model and DA model. This means that MT can not only 
learn speech patterns from mixed speech, but also can help DA to boost noise robustness.
We conjecture that this is because physical constraints are involved in MT training, which leads to a 
regularization effect that guides DA to learn real patterns of speech, rather than simply pursuing 
a low cost function. 

\begin{table}[h!]
  \caption{Performance in EER\% and Top-1 Accuracy (Acc\%) of different training strategies when detecting keywords from very noisy speech.}
  \vspace{-1.5mm}
  \label{tab:Noisy1-10}
  \centering
  \scalebox{0.8}{
  \begin{tabular}{lcccccc}
  \toprule
    \multirow{3}{*}{Test Set} & \multicolumn{2}{c}{Vanilla CNN} & \multicolumn{2}{c}{EfficientNet\_B0} & \multicolumn{2}{c}{EfficientNet\_B2} \\
                                \cmidrule(r){2-3}                 \cmidrule(r){4-5}                    \cmidrule(r){6-7}
                              &   EER   &  Top1 Acc                  &   EER   &  Top1 Acc                       &   EER    &    Top1 Acc   \\
        \cmidrule(r){1-1}      \cmidrule(r){2-3}                  \cmidrule(r){4-5}                      \cmidrule(r){6-7}
        Clean                 &   43.21 &   16.05               &   42.80 &  18.47                     &  43.11   &  18.59 \\
        DA                    &   28.55 &   38.65               &   25.38 &  43.03                     &  24.93   &  44.74 \\
        Mixup                 &   35.93 &   24.09               &   36.46 &  22.25                     &  35.74   &  23.52 \\
        Mixup (U)             &   35.53 &   25.46               &   35.85 &  23.93                     &  35.31   &  23.44 \\
        MT                    &   34.31 &   28.16               &   36.09 &  27.46                     &  34.83   &  28.53 \\
        MT (N)          &\textbf{27.59} &\textbf{43.58}   &\textbf{23.84} &\textbf{48.30}        &\textbf{22.78}  &\textbf{50.06} \\
  \bottomrule
  \end{tabular}}
\end{table}
\vspace{-1.8mm}
\subsection{Clean speech test}

In the last experiment, we tested whether MT and other robust training approaches have an impact on the performance of clean speech. 
The results are shown in Table~\ref{tab:clean}. Firstly, DA is not safe, especially with large models. 
Secondly, Mixup (U) is worse than the standard Mixup. The reason can be concluded as: the uniform distribution for the interpolation factor 
is just an ad-hoc hack to benefit mixed keyword detection, but loses the theoretical judgment on model regularization.
Finally, both Mixup and MT offer consistent performance gains, and their contributions are comparable. 
Note that these two techniques are based on different theoretical foundations. 
With large model EfficientNet-B0/B2, MT (N) obtained the best results, indicating that learning complicated patterns with large models may gain cheerful benefits in general.


\begin{table}[h!]
  \caption{Performance in EER\% and Top-1 Accuracy (Acc\%) of different training strategies when detecting keywords from clean speech.}
  \vspace{-1.5mm}
  \label{tab:clean}
  \centering
  \scalebox{0.8}{
  \begin{tabular}{lcccccc}
  \toprule
    \multirow{3}{*}{Test Set} & \multicolumn{2}{c}{Vanilla CNN} & \multicolumn{2}{c}{EfficientNet\_B0} & \multicolumn{2}{c}{EfficientNet\_B2} \\
                                \cmidrule(r){2-3}                 \cmidrule(r){4-5}                    \cmidrule(r){6-7}
                              &   EER   &  Top1 Acc                  &   EER   &  Top1 Acc                       &   EER    &    Top1 Acc   \\
        \cmidrule(r){1-1}      \cmidrule(r){2-3}                  \cmidrule(r){4-5}                      \cmidrule(r){6-7}
        Clean                 &   1.52          &   96.01               &   0.96           &  97.03             &  0.96             &  97.12 \\
        DA                    &   1.41          &   96.34               &   1.19           &  96.83             &  1.12             &  96.73 \\
        Mixup                 &\textbf{1.39}    &   96.67               &   0.96           &  97.79             &  0.91             &  97.87 \\
        Mixup (U)             &   1.75          &   96.16               &   1.02           &  97.61             &  1.09             &  97.87 \\
        MT                    &   1.50          &   \textbf{96.83}      &   \textbf{0.90}  &  97.79             &  1.14             &  97.96 \\
        MT (N)                &   1.66          &   96.71               &   \textbf{0.90}  &  \textbf{97.85}    &\textbf{0.85}      &  \textbf{98.10} \\
  \bottomrule
  \end{tabular}}
\vspace{-2.0mm}
\end{table}

\vspace{-1.8mm}
\section{Conclusions}

We proposed a mix training strategy to improve
the robustness of KWS models. The approach enforces 
models to detect keywords buried in interfering speech
or other keywords, rather than just minimizing the loss
function on clean data.
This is similar to human ears that must be `trained'
in noisy environments to gain sufficient sensitivity in complex 
conditions.

Extensive experiments were conducted, and the results showed
that mix training outperforms the popular data augmentation and
mixup training approaches by a large margin on tasks 
that detect keywords from very noisy and mixed speech.
Last but not least, mix training does not degrade performance
on clean speech, and in most cases, it can improve the performance.
Mix training is a general training strategy that can be applied to 
other speech process tasks and even to other superpositional signals,
which is our future work.



\newpage

\bibliographystyle{IEEEtran}
\bibliography{mybib}

\end{document}